\documentclass[preprint,prb,superscriptaddress,amssymb,amsmath, floatfix,citeautoscript]{revtex4-1}
\usepackage[pdftex]{graphicx}

\usepackage{amsmath,amssymb}

\begin{document}

\title{Cross-plane heat conduction in thin solid films}

\author{Chengyun Hua}

\author{Austin J. Minnich}%
 \email{aminnich@caltech.edu}
\affiliation{%
 Division of Engineering and Applied Science\\
 California Institute of Technology, Pasadena, California 91125,USA
}%

\date{\today}

\begin{abstract}

Cross-plane heat transport in thin films with thickness comparable to the phonon mean free paths is of both fundamental and practical interest. However, physical insight is difficult to obtain for the cross-plane geometry due to the challenge of solving the Boltzmann equation in a finite domain.  Here, we present a semi-analytical series expansion method to solve the transient, frequency-dependent Boltzmann transport equation that is valid from the diffusive to ballistic transport regimes and rigorously includes frequency-dependence of phonon properties. Further, our method is more than three orders of magnitude faster than prior numerical methods and provides a simple analytical expression for the thermal conductivity as a function of film thickness. Our result enables a more accurate understanding of heat conduction in thin films.

\end{abstract}

\maketitle 
\clearpage

\section{Introduction}

In the past two decades, thermal transport in thin solid films of thickness from tens of nanometers to micrometers has become a topic of considerable importance.\cite{Cahill2014} Such films occur in applications ranging from quantum well lasers to electronic devices.\cite{Chowdhury:2009,Su:2012a, Moore2014} For example, boundary scattering in these films leads to reduced thermal conductivity that results in the inefficient removal of heat in GaN transistors and LEDs.\cite{Balandin2011, Malen2012,Goodson2014} To address these and other problems, it is first necessary to understand the fundamental physics of heat conduction in micro-scale solid thin films.   

Heat transport in thin films with thickness comparable to the phonon mean free paths (MFPs) is governed by the Boltzmann transport equation (BTE), which is an integro-differential equation of time, real space and phase space. Due to its high dimensionality, it is in general very challenging to solve. For transport along the thin film, an analytical solution can be easily derived because the temperature gradient occurs along the infinite direction, simplifying the mathematics. Analytical solutions were derived for electron transport by Fuchs and Sondheimer with partially specular and partially diffuse boundary scattering.\cite{Fuchs1938, Sondheimer1952} Later, the Fuchs-Sondheimer solutions were extended to phonon thermal transport assuming an average phonon MFP, enabling the calculation of thermal conductivity as a function of the film thickness.\cite{Chen1993,Chen1997} Mazumder and Majumdar used a Monte-Carlo method to study the phonon transport along a silicon thin film including dispersion and polarization.\cite{Mazumder2001}

Heat conduction perpendicular to the thin film (cross-plane direction) is much more challenging. In order fields such as neutron transport and thermal radiation, solutions to the BTE for a slab geometry have been obtained using an invariant embedding method\cite{Chandrasekhar1950,BellmanBook}, an iterative method\cite{Case1967} and an eigenfunction expansion approach.\cite{Lii1972} For heat conduction, Majumdar numerically solved the gray phonon Boltzmann transport using a discrete-ordinate method by assuming that the two surfaces of the film were black phonon emitters.\cite{Majumdar1993} Later, Joshi and Majumdar developed an equation of phonon radiative transfer for both steady-state and transient cases, which showed the correct limiting behavior for both purely ballistic and diffusive transport.\cite{Joshi1993} Chen and Tien applied solutions from radiative heat transfer to calculate the thermal conductivity of a thin film attached to two thermal reservoirs.\cite{Chen1993} Chen obtained approximate analytical solutions of the BTE to study ballistic phonon transport in the cross-plane direction of superlattices and addressed the inconsistent use of temperature definition at the interfaces.\cite{Chen1998} Cross-plane heat conduction using a consistent temperature definition was then re-investigated by Chen and Zeng.\cite{Chen2001,Zeng2001}
 
Despite these extensive efforts to study transport in thin films based on the BTE, solutions for the cross-plane geometry are still only available with expensive numerical calculations. For example, no analogous Fuchs-Sondheimer formula for the in-plane thermal conductivity exists for the cross-plane direction. Further, most of the previous approaches assumed a single phonon MFP even though recent work has demonstrated that the transport properties of phonons in solids vary widely over the broad thermal spectrum.\cite{Esfarjani:2011,Broido2010} Incorporating frequency-dependent phonon properties with these prior numerical methods is extremely computationally expensive.

In this work, we present a semi-analytical solution of the frequency-dependent transient BTE using the method of degenerate kernels, also known as a series expansion method.\cite{IntegralEquation} Our approach that is valid from the diffusive to ballistic transport regimes, is capable of incorporating a variety of boundary conditions, and are more than three orders of magnitude faster than prior numerical approaches. Further, we obtain the equivalent of the Fuchs-Sondheimer analytical formula for the cross-plane thermal conductivity, enabling the cross-plane thermal conductivity of a thin film to be easily calculated.

\section{Theory}

\subsection{Method of degenerate kernels}\label{sec:method}
The one-dimensional (1D) frequency-dependent BTE for an isotropic crystal under the relaxation time approximation is given by:
\begin{equation}\label{eq:BTE_blackwall}
\frac{\partial g_{\omega}}{\partial t}+v_g\mu \frac{\partial g_{\omega}}{\partial x} = -\frac{ g_{\omega}-g_0(T)}{\tau_{\omega}}+\frac{Q_{\omega}}{4\pi}
\end{equation}
where $g_{\omega} = \hbar\omega(f_{\omega}(x,t,\theta)-f_0(T_0))$ is the desired deviational energy distribution function, $g_0(T)$ is the equilibrium deviational distribution function defined below, $Q_{\omega}$ is the spectral volumetric heat generation, $v_g$ is the phonon group velocity, and $\tau_{\omega}$ is the phonon relaxation time. Here, $x$ is the spatial variable, $t$ is the time, $\omega$ is the phonon frequency, $T$ is the temperature and $\mu = cos(\theta)$ is the directional cosine of the polar angle. The crystal is assumed to be isotropic.

Assuming a small temperature rise, $\Delta T = T - T_0$, relative to a reference temperature, $T_0$, the equilibrium deviational distribution is proportional to $\Delta T$, 
\begin{equation}
g_0(T) = \frac{1}{4\pi}\hbar \omega D(\omega) (f_{BE}(T) - f_{BE}(T_0)) \approx \frac{1}{4\pi}C_{\omega}\Delta T
\label{eq:BEDist_Linearized}
\end{equation}
 Here, $\hbar$ is the reduced Planck constant, $D(\omega)$ is the phonon density of states, $f_{BE}$ is the Bose-Einstein distribution, and $C_{\omega} = \hbar\omega D(\omega)\frac{\partial f_{BE}}{\partial T}$ is the mode specific heat. The volumetric heat capacity is then given by $C = \int_0^{\omega_m}C_{\omega}d\omega$ and the thermal conductivity $k = \int_0^{\omega_m}k_{\omega}d\omega$, where $k_{\omega} = \frac{1}{3} C_{\omega}v_{\omega} \Lambda_{\omega}$ and $\Lambda_{\omega} = \tau_{\omega}v_{\omega}$ is the phonon MFP. 

Both $g_{\omega}$ and $\Delta T$ are unknown. Therefore, to close the problem, energy conservation is used to relate $g_{\omega}$ to $\Delta T$, given by 
\begin{equation}
\int\int_0^{\omega_m} \left[\frac{g_{\omega}(x,t)}{\tau_{\omega}}-\frac{1}{4\pi}\frac{C_{\omega}}{\tau_{\omega}}\Delta T(x,t) \right]d\omega d\Omega = 0
\label{eq:EnergyConservation}
\end{equation}
where $\Omega$ is the solid angle in spherical coordinates and $\omega_m$ is the cut-off frequency. Note that summation over phonon branches is implied without an explicit summation sign whenever an integration over phonon frequency is performed. 

To solve this equation, we first transform it into an inhomogeneous first-order differential equation by applying a Fourier transform to the time variable, giving:
\begin{equation}\label{eq:BTE_TimeFourier}
i\eta \widetilde{g}_{\omega}+ v_g\mu \frac{ d \widetilde{g}_{\omega}}{dx} = -\frac{ \widetilde{g}_{\omega}}{\tau_{\omega}}+\frac{C_{\omega}}{\tau_{\omega}}\frac{\Delta \widetilde{T}}{4\pi}+\frac{\widetilde{Q}_{\omega}}{4\pi}
\end{equation}
where $\eta$ is the temporal frequency.  This equation has the general solution:
\begin{eqnarray}\label{eq:ODEsolution_gPlus}
\widetilde{g}^+_{\omega}(x) &=& P_{\omega}e^{-\frac{\gamma_{\omega}}{\mu}x}+\int_0^x\frac{C_{\omega}\Delta \widetilde{T}(x')+\widetilde{Q}_{\omega}(x')\tau_{\omega}}{4\pi\Lambda_{\omega}\mu}e^{\frac{\gamma_{\omega}}{\mu}(x'-x)}dx'\ (\mu \in (0, 1]) \\ \label{eq:ODEsolution_gMinus}
\widetilde{g}^-_{\omega}(x) &=& B_{\omega}e^{\frac{\gamma_{\omega}}{\mu}(L-x)}-\int_x^L\frac{C_{\omega}\Delta \widetilde{T}(x')+\widetilde{Q}_{\omega}(x')\tau_{\omega}}{4\pi\Lambda_{\omega}\mu}e^{\frac{\gamma_{\omega}}{\mu}(x'-x)}dx'\ (\mu \in [-1, 0])
\end{eqnarray}
where $\gamma_{\omega} = (1+i\eta\tau_{\omega})/\Lambda_{\omega}$, $L$ is the distance between the two walls, and $P_{\omega}$ and $B_{\omega}$ are the unknown coefficients determined by the boundary conditions. Here, $\widetilde{g}^+(x)$ indicates the forward-going phonons and $\widetilde{g}^-(x)$ the backward-going phonons. In this work, $\widetilde{g}^+(x)$ is specified at one of the two walls and $\widetilde{g}^-(x)$ is specified at the other. 

Let us assume that the two boundaries are nonblack but diffuse with wall temperature $\Delta T_1$ and $\Delta T_2$, respectively. The boundary conditions can be written as: 
\begin{eqnarray}\label{eq:diffuseBC_1}
\widetilde{g}^+_{\omega}(x = 0) &=& P_{\omega} =  \epsilon_1 \frac{C_{\omega}}{4\pi}\Delta T_1 + (1-\epsilon_1) \int_{-1}^0 \widetilde{g}^-_{\omega}(x=0,\mu)d\mu\\ 
\widetilde{g}^-_{\omega}(x = L) &=& B_{\omega} =\epsilon_2 \frac{C_{\omega}}{4\pi}\Delta T_2 +(1-\epsilon_2) \int_0^1 \widetilde{g}^+_{\omega}(x=L,\mu)d\mu \label{eq:diffuseBC_2}
\end{eqnarray}
where $\epsilon_1$ and $\epsilon_2$ are the emissivities of the hot and cold walls, respectively. When $\epsilon_1 = \epsilon_2 = 1$, the walls are black and we recover Dirichlet boundary conditions. Note that while we assume a thermal spectral distribution for the boundary condition, an arbitrary spectral profile can be specified by replacing the thermal distribution with the desired distribution. Equations (\ref{eq:diffuseBC_1}) \& (\ref{eq:diffuseBC_2}) are specific for diffuse boundary scattering; the specular case is presented in Appendix \ref{App:SpecularBoundaries}.

Applying the boundary conditions to Eqs.~(\ref{eq:ODEsolution_gPlus}) \& (\ref{eq:ODEsolution_gMinus}), we have
\begin{eqnarray}\label{eq:ODEsolution_g1}\nonumber
\widetilde{g}^+_{\omega}(x) &=& A_{1\omega}\frac{C_{\omega}}{4\pi} e^{-\frac{\gamma_{\omega}}{\mu}x}+e^{-\frac{\gamma_{\omega}}{\mu}x}\int_0^L\frac{C_{\omega}\Delta \widetilde{T}(x')+\widetilde{Q}_{\omega}(x')\tau_{\omega}}{4\pi\Lambda_{\omega}}\left[D_{\omega}E_{1}(\gamma_{\omega}(L-x'))+B_{1\omega}E_{1}(\gamma_{\omega}x')\right]dx'\\
&+&\int_0^x\frac{C_{\omega}\Delta \widetilde{T}(x')+\widetilde{Q}_{\omega}(x')\tau_{\omega}}{4\pi\Lambda_{\omega}\mu}e^{\frac{\gamma_{\omega}}{\mu}(x'-x)}dx'\ (\mu \in [0, 1]) \\  \nonumber
\widetilde{g}^-_{\omega}(x) &=& A_{2\omega}\frac{C_{\omega}}{4\pi} e^{-\frac{\gamma_{\omega}}{\mu}(L-x)}+e^{-\frac{\gamma_{\omega}}{\mu}(L-x)}\int_0^L\frac{C_{\omega}\Delta \widetilde{T}(x')+\widetilde{Q}_{\omega}(x')\tau_{\omega}}{4\pi\Lambda_{\omega}}\left[D_{\omega}E_{1}(\gamma_{\omega}x')+B_{2\omega}E_{1}(\gamma_{\omega}(L-x'))\right]dx'\\
&+&\int_x^L \frac{C_{\omega}\Delta \widetilde{T}(x')+\widetilde{Q}_{\omega}(x')\tau_{\omega}}{4\pi\Lambda_{\omega}\mu}e^{-\frac{\gamma_{\omega}}{\mu}(x'-x)}dx'\ (\mu \in [0, 1])
\label{eq:ODEsolution_g2}
\end{eqnarray}
where 
\begin{eqnarray}\nonumber
A_{1\omega} &=&\frac{\epsilon_1\Delta T_1+(1-\epsilon_1)\epsilon_2\Delta T_2 E_{2}(\gamma_{\omega}L)}{1-(1-\epsilon_1)(1-\epsilon_2)(E_{2}(\gamma_{\omega}L))^2},\  A_{2\omega} = \frac{\epsilon_2\Delta T_2+(1-\epsilon_2)\epsilon_1\Delta T_1 E_{2}(\gamma_{\omega}L)}{1-(1-\epsilon_1)(1-\epsilon_2)(E_{2}(\gamma_{\omega}L))^2}, \\ \nonumber
B_{1\omega} &=& \frac{1-\epsilon_1}{1-(1-\epsilon_1)(1-\epsilon_2)(E_{2}(\gamma_{\omega}L))^2},\  B_{2\omega} = \frac{1-\epsilon_2}{1-(1-\epsilon_1)(1-\epsilon_2)(E_{2}(\gamma_{\omega}L))^2}, \\ \nonumber
D_{\omega} &=& \frac{(1-\epsilon_1)(1-\epsilon_2)E_{2}(\gamma_{\omega}L)}{1-(1-\epsilon_1)(1-\epsilon_2)(E_{2}(\gamma_{\omega}L))^2}
\end{eqnarray}
and $E_n(x)$ is the exponential integral given by:\cite{GangBook}
\begin{equation}
E_{n}(x) = \int_0^1 \mu^{n-2}e^{-\frac{x}{\mu}}d\mu
\end{equation}
To close the problem, we plug Eqs.~(\ref{eq:ODEsolution_g1}) \& (\ref{eq:ODEsolution_g2}) into Eq.~(\ref{eq:EnergyConservation}) and obtain an integral equation for temperature as:
\begin{eqnarray}\nonumber
&&2 \int_0^{\omega_m} \frac{C_{\omega}}{\tau_{\omega}} d\omega \Delta \widetilde{T}(\widehat{x}) = \int_0^{\omega_m}\frac{C_{\omega}}{\tau_{\omega}}\left[A_{1\omega}E_{2}\left(\widehat{\gamma}_{\omega}\widehat{x}\right)+A_{2\omega}E_{2}(\widehat{\gamma}_{\omega}(1-\widehat{x}))\right]d\omega \\
&&+ \int_0^1 \int_0^{\omega_m} \widetilde{Q}_{\omega}(x') \frac{G_{\omega}(\widehat{x},\widehat{x}')}{\text{Kn}_{\omega}}d\omega d\widehat{x}' + \int_0^1 \Delta \widetilde{T}(\widehat{x}')\int_0^{\omega_m} \frac{C_{\omega}G_{\omega}(\widehat{x},\widehat{x}')}{\text{Kn}_{\omega}\tau_{\omega}}d\omega d\widehat{x}'
\label{eq:Temperature_blackwall}
\end{eqnarray}
where $\widehat{x} = x/L$,  Kn$_{\omega} = \Lambda_{\omega}/L$ is the Knudsen number, $\widehat{\gamma}_{\omega} = \frac{1+i\eta\tau_{\omega}}{\text{Kn}_{\omega}}$ and
\begin{eqnarray}\nonumber
G_{\omega}(\widehat{x},\widehat{x}') &=&  E_{2}(\widehat{\gamma}_{\omega}\widehat{x})\left[D_{\omega}E_{1}(\widehat{\gamma}_{\omega}(1-\widehat{x}'))+ B_{1\omega}E_{1}(\widehat{\gamma}_{\omega}\widehat{x}')\right] \\
&+& E_{2}(\widehat{\gamma}_{\omega}(1-\widehat{x}))\left[D_{\omega}E_{1}(\widehat{\gamma}_{\omega}\widehat{x}')+B_{1\omega}E_{1}(\widehat{\gamma}_{\omega}(1-\widehat{x}'))\right] + E_{1}(\widehat{\gamma}_{\omega}|\widehat{x}'-\widehat{x}|)
\end{eqnarray}
Equation (\ref{eq:Temperature_blackwall}) can be written in the form:
\begin{equation}\label{eq:Temperature_simplified}
\Delta T(\widehat{x}) = f(\widehat{x})+\int_0^1 K(\widehat{x},\widehat{x}') \Delta T(\widehat{x}') d\widehat{x}'
\end{equation}
where the kernel function $K(\widehat{x},\widehat{x}')$ is given by 
\begin{equation}\label{eq:FunctionK}
K(\widehat{x},\widehat{x}') = \frac{1}{2 \int_0^{\omega_m} \frac{C_{\omega}}{\tau_{\omega}} d\omega}\int_0^{\omega_m} \frac{C_{\omega}G_{\omega}(\widehat{x},\widehat{x}')}{\text{Kn}_{\omega}\tau_{\omega}}d\omega
\end{equation}
and the inhomogeneous function $f(\widehat{x})$ is given by
\begin{eqnarray}\label{eq:Functionf}\nonumber
f(\widehat{x}) &=& \frac{1}{2 \int_0^{\omega_m} \frac{C_{\omega}}{\tau_{\omega}} d\omega}\int_0^{\omega_m}\frac{C_{\omega}}{\tau_{\omega}}\left[A_{1\omega}E_{2}(\widehat{\gamma}_{\omega}\widehat{x}) + A_{2\omega}E_{2}(\widehat{\gamma}_{\omega}(1-\widehat{x}))\right]d\omega \\
&+& \frac{1}{2 \int_0^{\omega_m} \frac{C_{\omega}}{\tau_{\omega}} d\omega}\int_0^1 \int_0^{\omega_m} \widetilde{Q}_{\omega}(x') \frac{G_{\omega}(\widehat{x},\widehat{x}')}{\text{Kn}_{\omega}}d\omega d\widehat{x}'
\end{eqnarray}

From Eq.~(\ref{eq:Temperature_simplified}), we see that the governing equation is a Fredholm integral equation of the second kind. Previously, the gray version of Eq.~(\ref{eq:Temperature_blackwall}) that assumes average phonon properties has been solved numerically using finite differences.\cite{GangBook} While this approach does yield the solution, it requires the filling and inversion of a large, dense matrix, an expensive calculation even for the gray case. Considering frequency-dependence adds an additional integration to calculate each element of the matrix, dramatically increasing the computational cost. Additionally, care must be taken to account for a singularity point at $\widehat{x}' = \widehat{x}$ since $E_{1}(0) \rightarrow \infty$. Special treatment is required to treat this singularity point before discretizing the integral.  

Here, we will solve this equation using the method of degenerate kernels,\cite{IntegralEquation} which is much more efficient than the finite difference method and automatically accounts for the singularity point at $\widehat{x}' = \widehat{x}$.  This method is based on expanding all the functions in Eq.~(\ref{eq:Temperature_simplified}) in a Fourier series, then solving for the coefficients of the temperature profile. From the temperature $\Delta \widetilde{T}(\widehat{x})$, all other quantities such as the distribution and heat flux can be obtained. 

To apply this method, we first need to expand the inhomogeneous function $f(\widehat{x})$ and kernel $K(\widehat{x},\widehat{x}')$ with a Fourier series. This expansion is possible because both $f(\widehat{x})$ and $K(\widehat{x},\widehat{x}')$ are continuous and continuously differentiable on the relevant spatial domains of normalized length between $[0,1]$ and $[0,1]\times[0,1]$, respectively.\cite{IntegralEquation} All the necessary functions can be expanded using a linear combination of sines and cosines; however, a substantial simplification can be obtained by solving a symmetric problem in which the spatial domain is extended to include its mirror image by extending both $f(\widehat{x})$ and $K(\widehat{x},\widehat{x}')$ to [-1,1] and [-1,1] $\times$ [-1,1]. Because of the symmetry of this domain, all the coefficients for sine functions equal zero and the Fourier series for both functions reduces to a cosine expansion. $f(\widehat{x})$ is then approximated as
\begin{equation}\label{eq:f_expansion}
f_{(N)}(\widehat{x}) = \frac{1}{2}f_0 + \sum_{m = 1}^N f_m \text{cos}(m\pi \widehat{x})
\end{equation}
where $f_m = 2\int_0^1 f(\widehat{x})\text{cos}(m\pi \widehat{x})d\widehat{x}$.  The kernel $K(\widehat{x},\widehat{x}')$ can be represented by a degenerate double Fourier series, given by
\begin{equation}\label{eq:k_expansion}
K_{(N)}(\widehat{x},\widehat{x}') = \frac{1}{4} k_{00} + \frac{1}{2}\sum_{m=1}^N k_{m0} \text{cos}(m\pi \widehat{x})+ \frac{1}{2}\sum_{n=1}^N k_{0n} \text{cos}(n\pi \widehat{x}')+\sum_{m=1}^N \sum_{n=1}^N k_{mn}\text{cos}(m\pi\widehat{x})\text{cos}(n\pi\widehat{x}')
\end{equation}
where 
\begin{equation}\label{eq:k_fouriercoeff}
k_{mn} = 4\int_0^1 \int_0^1 K(\widehat{x},\widehat{x}')\text{cos}(m\pi \widehat{x})\text{cos}(n\pi \widehat{x}') d\widehat{x}d\widehat{x}'
\end{equation} 
Moreover, the convergence and completeness theorem of cosine functions guarantees that $K_{(N)}(\widehat{x},\widehat{x}')$ and $f_{(N)}(\widehat{x})$ converge to $K(\widehat{x},\widehat{x}')$ and $f(\widehat{x})$ as $N \rightarrow \infty$.\cite{PDE}

Inserting Eqs.~(\ref{eq:f_expansion}) \& (\ref{eq:k_expansion}) into Eq.~(\ref{eq:Temperature_simplified}), we then obtain the following integral equation
\begin{eqnarray}\nonumber
\sum_{m=0}^{N} x_m \text{cos}(m\pi\widehat{x})  &=&\frac{1}{2}f_0+ \sum_{m=0}^{N} f_m \text{cos}(m\pi\widehat{x}) + \int_0^1 \sum_{n=0}^{N} x_m \text{cos}(n\pi\widehat{x}')\left[\frac{1}{4} k_{00} + \frac{1}{2}\sum_{m=1}^N k_{m0} \text{cos}(m\pi \widehat{x})\right. \\
&& \left. + \frac{1}{2}\sum_{n=1}^N k_{0n} \text{cos}(n\pi \widehat{x}')+\sum_{m=1}^N \sum_{n=1}^N k_{mn}\text{cos}(m\pi\widehat{x})\text{cos}(n\pi\widehat{x}')\right]d\widehat{x}'
\label{eq:Temperature_FourierSeries}
\end{eqnarray}
where $x_m$ are the desired but unknown Fourier coefficients of $\Delta \widetilde{T}(\widehat{x})$.

Using the orthogonality of cos$(n\pi\widehat{x})$ on $[0,1]$ gives a simpler form of Eq.~(\ref{eq:Temperature_FourierSeries}):
\begin{equation}
 \sum_{m=0}^{N} x_m \text{cos}(m\pi\widehat{x})  = \frac{1}{2}f_0+ \sum_{m=0}^{N} f_m \text{cos}(m\pi\widehat{x}) +\frac{1}{4}\sum_{m=0}^N k_{m0}x_m + \frac{1}{2}\sum_{m=1}^N\sum_{n=1}^N k_{mn}x_n\text{cos}(m\pi\widehat{x})
\end{equation}

Grouping the terms with the same index $m$ in cosine, a system of linear equations of $x_m$ can be obtained as:
\begin{equation}\label{eq:LinearMatrix}
\Bar{\Bar{A}} \Bar{x} = \Bar{f}
\end{equation}
where $\Bar{x}$ is the vector of unknown coefficient $x_m$ and $\Bar{f}$ is the vector of $f_m$ in Eq.~(\ref{eq:f_expansion}). The matrix $\Bar{\Bar{A}}$ contains elements $A_{00} = 2-\frac{k_{00}}{2}$, $A_{mm} =1-\frac{k_{m0}}{2}$ for $m\neq 0$, and $A_{mn} = \frac{k_{mn}}{2}$ for $m \neq n$. Expressions of the elements in $\Bar{\Bar{A}}$ can be obtained analytically for the specific kernel here and are given in Appendix \ref{App:FourierCoefficients} for the steady-state heat conduction with diffuse walls. Since there is no row or column in $\Bar{\Bar{A}}$ that is all zeros or a constant multiple of another row or column, it is always guaranteed that $\Bar{\Bar{A}}$ is non-singular and its inverse exists. 

Solving the matrix system yields $x_m$ and thus the temperature $\Delta \widetilde{T}(\widehat{x})$. $\widetilde{g}^+_{\omega}(x)$ and $\widetilde{g}^-_{\omega}(x)$ can be obtained from $\Delta \widetilde{T}(\widehat{x})$ using Eqs.~(\ref{eq:ODEsolution_g1}) \& (\ref{eq:ODEsolution_g2}). Finally, the spectral heat is given by:
\begin{equation}
q_{\omega}(x) = \int_{-1}^1 g_{\omega} v_{\omega}\mu d\mu =\int_0^1 g^{+}_{\omega} v_{\omega}\mu d\mu - \int_0^1 g^{-}_{\omega}v_{\omega}\mu d\mu
\end{equation}
thereby closing the problem.

\subsection{Summary of the method}

We now summarize the key steps to implement the method of degenerate kernels.The first step is to determine the appropriate boundary conditions for the problem and compute the constants in Eqs.~(\ref{eq:ODEsolution_gPlus}) \& (\ref{eq:ODEsolution_gMinus}). Subsequently, the kernel function $K(\widehat{x},\widehat{x}')$ and the inhomogenous function $f(\widehat{x})$ can be obtained from Eq.~(\ref{eq:EnergyConservation}), and their Fourier coefficients can be computed using Eqs.~(\ref{eq:f_expansion}) \& (\ref{eq:k_expansion}). The elements in $\Bar{\Bar{A}}$ correspond to the Fourier coefficients of  kernel function $K(\widehat{x},\widehat{x}')$, and $\bar{f}$ is a vector of the Fourier coefficients of the inhomogeneous part of Eq.~(\ref{eq:Temperature_blackwall}). We emphasize that analytic expressions for all of these elements exist can be obtained; examples of these coefficients for steady heat conduction with non-black, diffuse boundaries are given in Appendix \ref{App:FourierCoefficients}. Once $\Bar{\Bar{A}}$ and $\bar{f}$ are obtained, Eq.~(\ref{eq:LinearMatrix}) is solved by standard matrix methods to yield the coefficients $x_m$. Finally, $\Delta \widetilde{T}(\widehat{x})$ is given by $\sum_{m=0}^{N} x_m \text{cos}(m\pi\widehat{x})$.

\subsection{Computational efficiency of the method}

Since both $K_{(N)}(\widehat{x},\widehat{x}')$ and $f_{(N)}(\widehat{x})$ converge to $K(\widehat{x},\widehat{x}')$ and $f(\widehat{x})$ as $1/N^2$, only a few terms of expansion are required for accurate calculations. In practice, we find that only 20 terms are necessary before the calculation converges, meaning the required matrix is only $20 \times 20$. Compared to the traditional discretization method that requires a matrix on the order of $1000\times 1000$ before convergence is achieved, our approach is at least 3 orders of magnitude faster. 

\section{Application}\label{sec:applications}

To illustrate the method, we consider steady-state heat conduction between two walls that are either black or non-black. In the former case, both wall emissivities $\epsilon_1$ and $\epsilon_2$ equal 1 while in the latter case they are set to 0.5. Assuming steady state and no heat generation inside the domain, $\eta = 0$, and $Q_{\omega} = 0$. The Fourier coefficients of $K(\widehat{x},\widehat{x}')$ and $\bar{f}$ for these two specific cases are given in Appendix \ref{App:FourierCoefficients}.   

We perform our calculations for crystalline silicon, using the experimental dispersion in the [100] direction and assuming the crystals are isotropic. The numerical details concerning the dispersion and relaxation times are given in Ref. 27.\nocite{Minnich2011}

\begin{figure*}[t!]
\centering
\includegraphics[scale = 0.4]{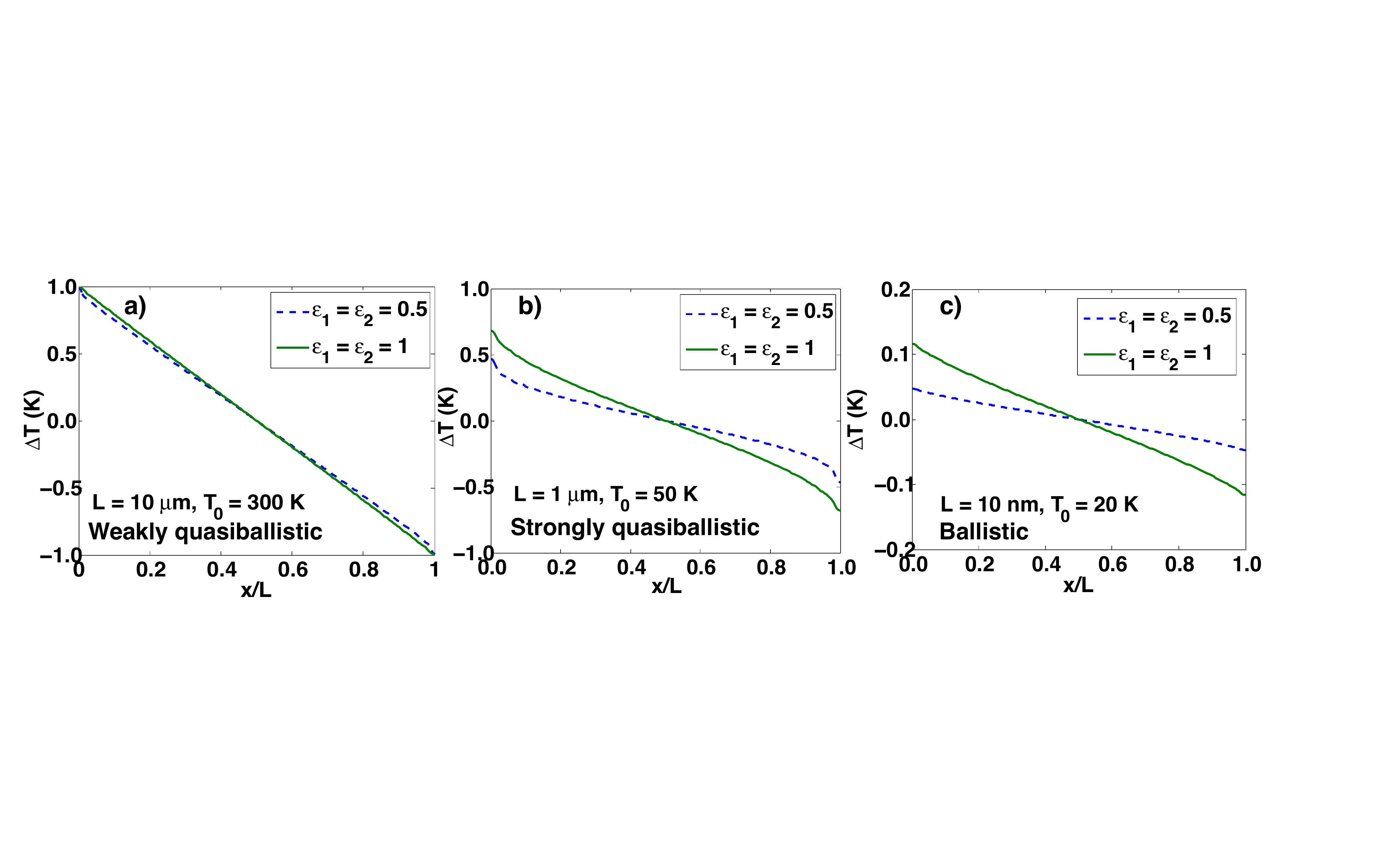}
\caption{Temperature distribution $\Delta T(\widehat{x})$ for a planar slab with black walls (solid lines) and nonblack walls (dashed lines) in the (a) weakly quasiballistic regime, (b) strongly quasiballistic regime, and (c) ballistic regime. As Kn$_{\omega}$ increases, temperature slip at the boundaries grows larger.}
\label{fig:Temperature}
\end{figure*}

\subsection{Temperature distribution}

We first calculate the deviational temperature distribution $\Delta T(\widehat{x})$ for different film thickness at different equilibrium temperatures as shown in Fig.~\ref{fig:Temperature} while keeping $|\Delta T1| = |\Delta T_2| = 1$ K. When phonon MFPs are much smaller than the film thickness and there are sufficient scattering events in the domain, as occurs for silicon at room temperature as $L \rightarrow \infty$, we recover the diffusion limit for both the black and non-black cases, which is a linear line between $\Delta T_1$ and $\Delta T_2$. 

As we decrease the film thickness such that it is comparable to the phonon MFPs, the transport becomes quasiballistic. Hua and Minnich\cite{Hua2014} have further divided this regime into weakly and strongly quasiballistic regimes that are distinguished by the portion of the phonon spectrum that is ballistic. The weakly quasiballistic regime is characterized by ballistic transport of low frequency, low heat capacity modes that affect thermal conductivity but not the temperature profile, while in the strongly quasiballistic regime high heat capacity modes are also ballistic. In the present steady state case, the two regimes are distinguished by the magnitude of the Knudsen number Kn$_{\omega}$; for example, when Kn$_{\omega} \sim O(10^{-4})$ or less, then the transport is diffusive.

In the weakly quasiballistic regime, Kn$_{\omega} \sim O(10^{-2}$), the temperature profile remains linear with negligible temperature slip at the boundaries as in the diffusion regime. While it is difficult to distinguish the diffusion and weakly quasiballistic regimes by the temperature profile, we will show in the next section that heat flux is substantially different between the two regimes. Note that when the walls are black, the temperature profile is identical to the diffusion case, but when the wall is nonblack, slight deviations occur as in Fig.~\ref{fig:Temperature}(a). These deviations occur because not all the phonons leaving the walls are at the boundary temperature; some phonons are at the temperature of the opposite wall and are reflected without thermalizing. 

In the strongly quasiballistic regime where Kn$_{\omega} \sim 1$, the temperature distribution is no longer linear and the modified Fourier law breaks down. We observe temperature slip at the two boundaries shown in Fig.~\ref{fig:Temperature}(b). Physically, temperature slip occurs because the emitted phonon temperatures, $T_1$ and $T_2$ do not represent the local energy density in the solid due to lack of scattering events. For nonblack wall case, the temperature discontinuities at the two boundaries are bigger than the black case. This difference occurs because part of the phonon distribution consists of diffusely reflected phonons that have not thermalized at the boundary, leading to a bigger difference between the local and emitted energy density distributions than in the black case and hence a larger temperature slip. 

As Kn$_{\omega} \gg 1$ as shown in Fig.~\ref{fig:Temperature}(c), we approach the ballistic limit of phonon transport. In this limit, phonons propagating from one wall to the other do not interact with phonons emitted from the other wall due to the complete absence of scattering. As Kn$_{\omega} \rightarrow \infty$, the phonon temperature throughout the domain approaches $(T_1+T_2)/2$, the average of the phonon temperatures emitted from each wall.

\begin{figure*}[t!]
\centering
\includegraphics[scale = 0.55]{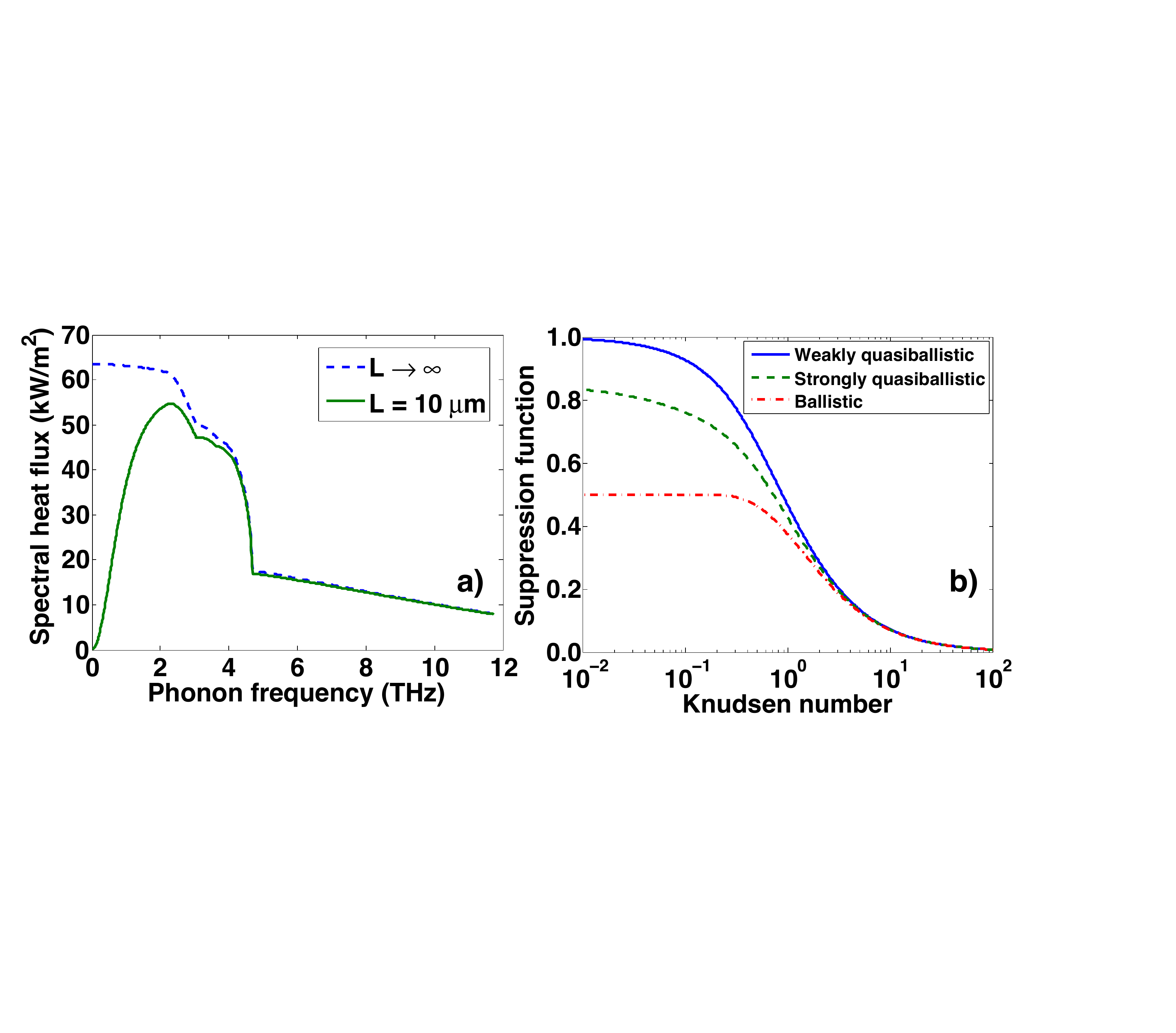}
\caption{(a) Spatially averaged spectral heat flux versus phonon frequency between two black walls at 301 K and at 299 K in the diffusive limit (dashed line) and in the weakly quasiballistic regime (solid line). For silicon at room temperature, the onset of the weakly quasiballistic transport starts at $L = 10$ $\mu m$. (b) Suppression function versus Knudsen number in the weakly quasiballistic regime (solid line), strongly quasiballistic regime (dashed line) and ballistic limit (dash-dotted line). In general, the suppression function depends not only on Knudsen number but also on the temperature distribution for a given thickness. }
\label{fig:Suppression}
\end{figure*}

\subsection{Heat flux and suppression function}

Next, we seek to understand how the thickness affects which phonons conduct heat in each regime. For simplicity, we focus on the black case.  From our model, we can calculate the spatially averaged spectral heat flux that is integrated over the domain, defined as:
\begin{eqnarray}\nonumber
&& \int_0^1 q_{\omega}(\widehat{x})d\widehat{x} = \frac{1}{L}\int_0^L q_{\omega}(x)dx =\left(\frac{\Delta T_1 - \Delta T_2}{2}\right)\left[\frac{1}{3}C_{\omega}v_{\omega}\text{Kn}_{\omega}-C_{\omega}v_{\omega}\text{Kn}_{\omega} E_{4}\left(\frac{1}{\text{Kn}_{\omega}}\right)\right] \\  \label{eq:AveragedHeatFlux} 
&&+ \frac{C_{\omega}v_{\omega}}{2\text{Kn}_{\omega}}\left[\int_0^1 \int_0^{\widehat{x}} \Delta T(\widehat{x}')E_2\left(\frac{|\widehat{x}'-\widehat{x}|}{\text{Kn}_{\omega}}\right) d\widehat{x}' d\widehat{x}-\int_0^1 \int_{\widehat{x}}^1 \Delta T(\widehat{x}')E_2\left(\frac{|\widehat{x}'-\widehat{x}|}{\text{Kn}_{\omega}}\right) d\widehat{x}' d\widehat{x}\right]
\end{eqnarray}
Once $x_m$ is solved from Eq.~(\ref{eq:Temperature_FourierSeries}), we can insert the Fourier series of $\Delta T(x)$ into Eq.~(\ref{eq:AveragedHeatFlux}), which leads to 
\begin{eqnarray}\nonumber
\int_0^1 q_{\omega}(\widehat{x})d\widehat{x} &=& \left[ \left(\frac{\Delta T_1 - \Delta T_2}{2}\right)\frac{1}{3}C_{\omega}v_{\omega}\text{Kn}_{\omega}-C_{\omega}v_{\omega}\text{Kn}_{\omega}  E_{4}\left(\frac{1}{\text{Kn}_{\omega}}\right) \right]\\
&+& \frac{C_{\omega}v_{\omega}}{2\text{Kn}_{\omega}}\sum_{m=1}^{\infty}x_m [1-(-1)^m]\int_0^1 \frac{(\text{Kn}_{\omega}\mu)^2\left(1+e^{-\frac{1}{\text{Kn}_{\omega}\mu}}\right)}{1+(\text{Kn}_{\omega}\mu)^2(m\pi)^2}d\mu
\end{eqnarray}
According to Fourier's law, the integrated heat flux is given by
\begin{equation}\label{eq:FourierHF}
\int_0^1 q_{\omega}^{f}(\widehat{x})d\widehat{x} = \frac{1}{3}C_{\omega}v_{\omega}\text{Kn}_{\omega}(\Delta T_1-\Delta T_2)
\end{equation}
The heat suppression function is defined as the ratio of the BTE and Fourier's heat flux\cite{Minnich2012}, given as
\begin{equation}\label{eq:suppression}
S(\text{Kn}_{\omega}, L) = \frac{1}{2}-\frac{3}{2}  E_{4}\left(\frac{1}{\text{Kn}_{\omega}}\right) +\frac{3}{2} \sum_{m=1}^{\infty}x_m [1-(-1)^m]\int_0^1 \frac{\mu^2\left(1+e^{-\frac{1}{\text{Kn}_{\omega}\mu}}\right)}{1+(\text{Kn}_{\omega}\mu)^2(m\pi)^2}d\mu
\end{equation}
Note that the suppression function in general not only depends on Kn$_{\omega}$ but also is a function of geometry through $x_m$. The reduced or apparent thermal conductivity at a given domain thickness $L$ is then given by:
\begin{equation}\label{eq:Reducedk}
k(L) = \int_0^{\omega_m}\frac{1}{3}C_{\omega}v_{\omega}\Lambda_{\omega}S(\text{Kn}_{\omega}, L)d\omega
\end{equation}
This formula is analogous to the Fuch-Sondheimer equation for transport along thin films and allows the simple evaluation of the cross-plane thermal conductivity once $x_m$ are known. 

Figure \ref{fig:Suppression}(a) shows the computed spectral heat flux from Eq.~(\ref{eq:FourierHF}) in the diffusive limit ($L\rightarrow \infty$) and in the weakly quasiballistic regime ($L = 10\ \mu m$). We note that even though the temperature distributions are nearly identical for these two regimes as shown in Fig.~\ref{fig:Temperature}(a), the heat carried by low frequency phonons in the weakly quasiballistic regime is much smaller than that in the diffusive limit, leading to a smaller effective thermal conductivity. 

While Eq.~(\ref{eq:suppression}) enables the calculation of the cross-plane thermal conductivity of a slab, it requires knowledge of the temperature profile. A more useful result would be a suppression function that depends only on the Knudsen number as is available for in-plane heat conduction with the Fuchs-Sondheimer formula.\cite{Fuchs1938, Sondheimer1952} To obtain this result, we  derive simplifications to the suppression function, Eq.~(\ref{eq:suppression}), in the weakly quasiballistic limit and completely ballistic limits. In the weakly quasiballistic regime, the temperature distribution is still linear, allowing us to simplify Eq.~(\ref{eq:suppression}) by inserting the linear temperature distribution. Doing so leads to the weak suppression function:
\begin{equation}\label{eq:suppression_weakly}
S_{\text{weak}}(\text{Kn}_{\omega}) = 1 + 3\text{Kn}_{\omega}\left[ E_{5}\left(\frac{1}{\text{Kn}_{\omega}}\right)-\frac{1}{4}\right]
\end{equation} 

On the order hand, in the ballistic limit, $\Delta T(\widehat{x}) = 0$ everywhere in the domain, which leads to a ballistic-limit suppression function,
\begin{equation}
S_{\text{ballistic}}(\text{Kn}_{\omega})= \frac{1}{2}-\frac{3}{2}E_{4}\left(\frac{1}{\text{Kn}_{\omega}}\right)
\end{equation}

Both of these equations depend only on Knudsen number and thus can be directly applied without explicitly solving for the temperature profile. Between these two regimes is the strongly quasiballistic regime, where a full expression given by Eq.~(\ref{eq:suppression}) is necessary. Figure \ref{fig:Suppression}(b) shows the suppression functions as a function of Kn$_{\omega}$ in these three regimes. Note that in the strongly quasiballistic regime, the suppression function depends not just on Knudsen number but also on material properties. We obtained these results for silicon with an equilibrium temperature at 50 K and a slab thickness of 1 $\mu$m.

One important observation from Fig.~\ref{fig:Suppression}(b) is that the suppression functions in the different regimes converge to the same curve at large Kn$_{\omega}$. Also note that as the slab thickness decreases, the Knudsen number of a phonon with a particular MFP becomes larger. Therefore, in the limit of very small distance between the boundaries, the only important portion of the suppression function is at large values of Knudsen number exceeding Kn$_{\omega} = 1$ because phonons possess a minimum MFP. This observation suggests that for practical purposes the weak suppression can be used even outside the range in which it is strictly valid with good accuracy. This simplification is very desirable because the weak suppression function only depends on the Knudsen number and thus can be applied without any knowledge of other material properties, unlike in the strongly quasiballistic regime. 

To investigate the accuracy of this approximation, we perform a reconstruction procedure developed by Minnich\cite{Minnich2012} to recover the MFP spectrum from thermal conductivity data as a function of slab thickness using both full and weak suppression functions. We follow the exact procedures of the numerical method as described in Ref.~29. Briefly, we synthesize effective thermal conductivities numerically using Eq.~(\ref{eq:Reducedk}). Using these effective thermal conductivities and our knowledge of suppression function, we use convex optimization to solve for the MFP spectrum. In the full suppression function case, each slab thickness has its own suppression function given by Eq.~(\ref{eq:suppression}) while in the weak case Eq.~(\ref{eq:suppression_weakly}) is used for all slab thicknesses.

\begin{figure*}[t!]
\centering
\includegraphics[scale = 0.55]{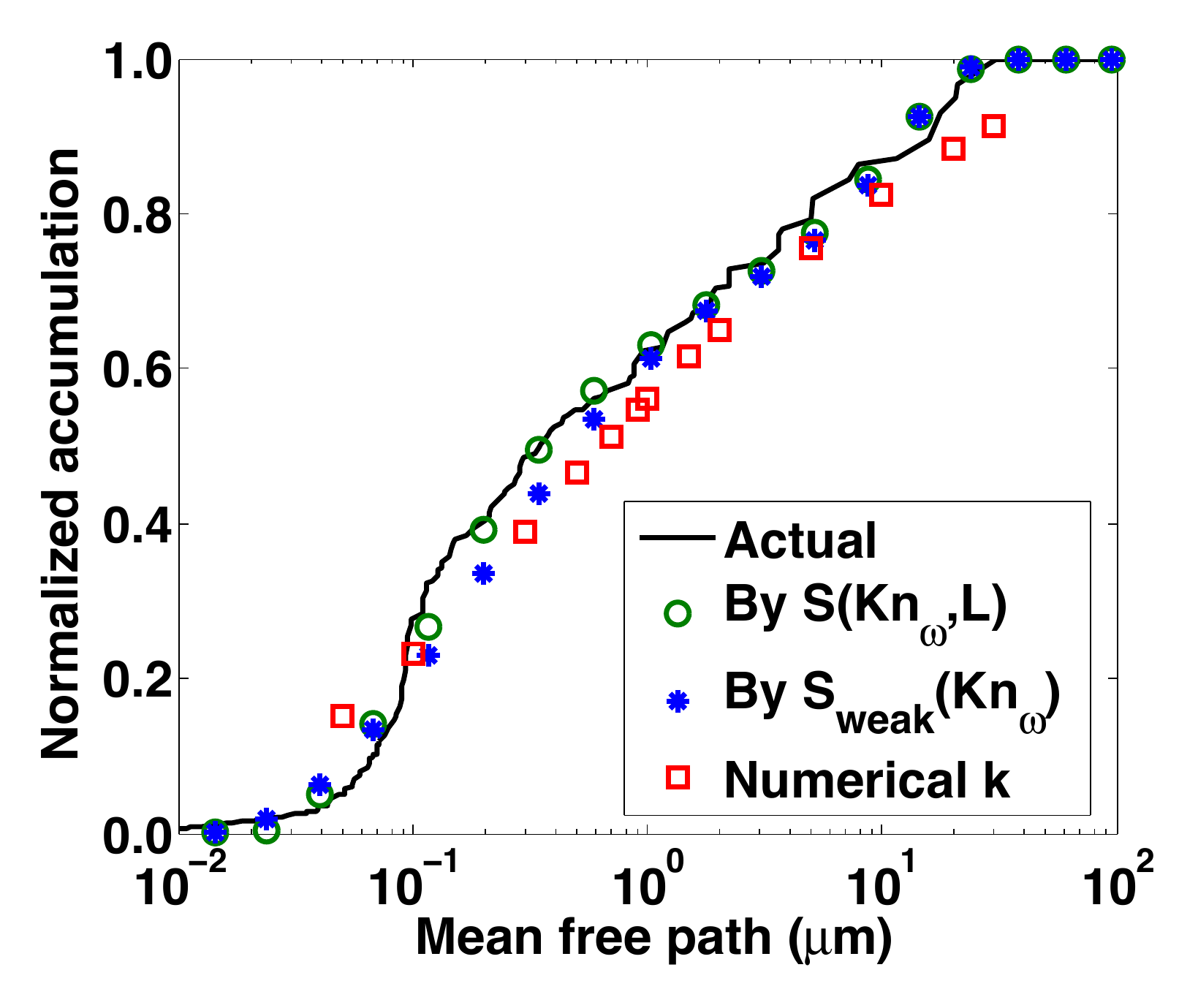}
\caption{Example MFP reconstructions for silicon at 300 K using numerically simulated data. Plotted are the analytical MFP distribution (solid line), the numerical apparent thermal conductivities (squares), and the reconstructed MFP distribution by the full suppression function (circles) and by the weak suppression function (stars). The $x$ axis corresponds to the MFP for the distributions and to the film thickness for the thermal conductivity data. Both the weak and full suppression functions yield satisfactory MFP reconstruction results.}
\label{fig:Reconstruction}
\end{figure*}

As shown in Fig.~\ref{fig:Reconstruction}, both the weak and full suppression functions yield satisfactory results. Even though the smallest thickness we consider here is 50 nm, close to the ballistic regime, the weak suppression function still gives a decent prediction over the whole MFP spectrum, with a maximum of 15 \% deviation from the actual MFP spectrum. For practical purposes, this deviation is comparable to uncertainties in experimental measurements and therefore the weak suppression function can be used as an excellent approximation in the reconstruction procedure.  This result demonstrates that length-dependent thermal conductivity measurements like those recently reported for SiGe nanowires\cite{Hsiao2013} and graphene ribbons\cite{Xu2014} can be used to reconstruct the full MFP spectrum rather than only an average MFP. We perform an investigation of our approach for this purpose in a separate article.

\section{Summary}

We have presented a series expansion method to solve the one-dimensional, transient frequency-dependent BTE in a finite domain and demonstrated its capability to describe cross-heat conduction in thin films. Our solution is valid from the diffusive to ballistic regimes with a variety of boundary conditions, rigorously includes frequency dependence, and is more than three orders of magnitude faster than prior numerical approaches. We have also developed a simple analytical expression for thermal conductivity, analogous to the Fuchs-Sondheimer equation for in-plane transport, than enables the simple calculation of the cross-plane thermal conductivity as a function of film thickness. Our work will enable a better understanding of cross-plane heat conduction in thin films.

\section{Acknowledgement}

This work was sponsored in part by Robert Bosch LLC through Bosch Energy Research Network Grant no. 13.01.CC11, by the National Science Foundation under Grant no. CBET CAREER 1254213, and by Boeing under the Boeing-Caltech Strategic Research \& Development Relationship Agreement.

\clearpage
\bibliographystyle{is-unsrt}
\bibliography{Myrefs}

\clearpage
\appendix
\section{Specular boundaries}\label{App:SpecularBoundaries}

Here, we consider the two boundaries to be nonblack but specular with wall temperature $\Delta T_1$ and $\Delta T_2$, respectively. The boundary conditions can be written as:
\begin{eqnarray}\label{eq:specularBC_1}
\widetilde{g}^+_{\omega}(x = 0,\mu) &=& P_{\omega} =  \epsilon_1 \frac{C_{\omega}}{4\pi}\Delta T_1 + (1-\epsilon_1) \widetilde{g}^-_{\omega}(x=0,-\mu)\\ 
\widetilde{g}^-_{\omega}(x = L,\mu) &=& B_{\omega} =\epsilon_2 \frac{C_{\omega}}{4\pi}\Delta T_2 +(1-\epsilon_2) \widetilde{g}^+_{\omega}(x=L,-\mu),\label{eq:specularBC_2}
\end{eqnarray}
Applying the boundary conditions to Eqs.~(\ref{eq:ODEsolution_gPlus}) \& (\ref{eq:ODEsolution_gMinus}), we have
\begin{eqnarray}\label{eq:g1_specular}\nonumber
\widetilde{g}^+_{\omega}(x) &=& F_1 \Delta T_1 \frac{C_{\omega}}{4\pi}e^{-\frac{\gamma_{\omega}}{\mu}x} + (1-\epsilon_1)F_2\Delta T_2  \frac{C_{\omega}}{4\pi}e^{-\frac{\gamma_{\omega}}{\mu}(L+x)}\\ \nonumber
&+& (1-\epsilon_1)F_2 \int_0^L \frac{C_{\omega}\Delta \widetilde{T}(x')+\widetilde{Q}_{\omega}(x')\tau_{\omega}}{4\pi\Lambda_{\omega}\mu}e^{-\frac{\gamma_{\omega}}{\mu}(x'+x)}dx'\\
&+& \int_0^x\frac{C_{\omega}\Delta \widetilde{T}(x')+\widetilde{Q}_{\omega}(x')\tau_{\omega}}{4\pi\Lambda_{\omega}\mu}e^{\frac{\gamma_{\omega}}{\mu}(x'-x)}dx'\ (\mu \in [0, 1]), \\  \nonumber
\widetilde{g}^-_{\omega}(x) &=& F_2\Delta T_2 \frac{C_{\omega}}{4\pi}e^{-\frac{\gamma_{\omega}}{\mu}(L-x)} + (1-\epsilon_2)F_1\Delta T_1 \frac{C_{\omega}}{4\pi}e^{-\frac{\gamma_{\omega}}{\mu}(2L-x)} \\ \nonumber
&+& (1-\epsilon_2)F_1\int_0^L \frac{C_{\omega}\Delta \widetilde{T}(x')+\widetilde{Q}_{\omega}(x')\tau_{\omega}}{4\pi\Lambda_{\omega}\mu}e^{-\frac{\gamma_{\omega}}{\mu}(2L-x'-x)}dx'\\
&+&\int_x^L \frac{C_{\omega}\Delta \widetilde{T}(x')+\widetilde{Q}_{\omega}(x')\tau_{\omega}}{4\pi\Lambda_{\omega}\mu}e^{-\frac{\gamma_{\omega}}{\mu}(x'-x)}dx'\ (\mu \in [0, 1]),
\label{eq:g2_specular}
\end{eqnarray}
where $F_1 = \frac{\epsilon_1}{\epsilon_1+\epsilon_2-\epsilon_1\epsilon_2}$ and $ F_2 = \frac{\epsilon_2}{\epsilon_1+\epsilon_2-\epsilon_1\epsilon_2}$.

To close the problem, we insert Eqs.~(\ref{eq:g1_specular}) \& (\ref{eq:g2_specular}) into Eq.~(\ref{eq:EnergyConservation}) and nondimensionalize $x$ by $L$. We then derive an integral equation for temperature for the specular boundary conditions, given by
\begin{eqnarray}\nonumber
2 \int_0^{\omega_m} \frac{C_{\omega}}{\tau_{\omega}} d\omega \Delta \widetilde{T}(\widehat{x}) &=& \int_0^{\omega_m}\frac{C_{\omega}}{\tau_{\omega}}H_{\omega}(\widehat{x})d\omega + \int_0^1 \int_0^{\omega_m} \widetilde{Q}_{\omega}(x') \frac{G_{\omega}(\widehat{x},\widehat{x}')}{\text{Kn}_{\omega}}d\omega d\widehat{x}'\\
&+& \int_0^1 \Delta \widetilde{T}(\widehat{x}')\int_0^{\omega_m} \frac{C_{\omega}G_{\omega}(\widehat{x},\widehat{x}')}{\text{Kn}_{\omega}\tau_{\omega}}d\omega d\widehat{x}',
\label{eq:Temperature_blackwall_specular}
\end{eqnarray}
where $\widehat{x} = x/L$,  Kn$_{\omega} = \Lambda_{\omega}/L$ is the Knudsen number, $\widehat{\gamma}_{\omega} = \frac{1+i\eta\tau_{\omega}}{\text{Kn}_{\omega}}$ and
\begin{eqnarray}\nonumber
H_{\omega}(\widehat{x}) &=& F_1\Delta T_1  E_2(\widehat{\gamma}_{\omega}\widehat{x})+F_2\Delta T_2 E_2(\widehat{\gamma}_{\omega}(1-\widehat{x}))\\
&+& (1-\epsilon_1)F_2\Delta T_2 E_2(\widehat{\gamma}_{\omega}(1+\widehat{x}))+(1-\epsilon_2)F_1\Delta T_1 E_2(\widehat{\gamma}_{\omega}(2-\widehat{x}))
\end{eqnarray}
and 
\begin{equation}
G_{\omega}(\widehat{x},\widehat{x}')= (1-\epsilon_1)F_2 E_1(\widehat{\gamma}_{\omega}(\widehat{x}+\widehat{x}'))+(1-\epsilon_2)F_1 E_1(\widehat{\gamma}_{\omega}(2-\widehat{x}-\widehat{x}'))+ E_1(\widehat{\gamma}_{\omega}|\widehat{x}-\widehat{x}'|).
\end{equation}
In this case, the inhomogeneous function becomes
\begin{equation}
f(\widehat{x}) = \frac{1}{2 \int_0^{\omega_m} \frac{C_{\omega}}{\tau_{\omega}} d\omega}\left[\int_0^{\omega_m}\frac{C_{\omega}}{\tau_{\omega}}H_{\omega}(\widehat{x})d\omega + \int_0^1 \int_0^{\omega_m} \widetilde{Q}_{\omega}(x') \frac{G_{\omega}(\widehat{x},\widehat{x}')}{\text{Kn}_{\omega}}d\omega d\widehat{x}'\right]
\end{equation}
and the kernel function becomes
\begin{equation}
K(\widehat{x},\widehat{x}') = \frac{1}{2 \int_0^{\omega_m} \frac{C_{\omega}}{\tau_{\omega}} d\omega}\int_0^{\omega_m} \frac{C_{\omega}G_{\omega}(\widehat{x},\widehat{x}')}{\text{Kn}_{\omega}\tau_{\omega}}d\omega.
\end{equation}

With these results, the problem can be solved by following the same procedures described in Sec.\ref{sec:method} are followed to formulate a linear system of equations. The solution of this system then yields the temperature Fourier coefficients. 

\section{Fourier coefficients}\label{App:FourierCoefficients}

For steady-state heat conduction between two non-black walls as studied in Sec.~\ref{sec:applications}, the inhomogeneous function becomes
\begin{equation}
f(\widehat{x}) = \frac{1}{2 \int_0^{\omega_m} \frac{C_{\omega}}{\tau_{\omega}} d\omega}\int_0^{\omega_m}\frac{C_{\omega}}{\tau_{\omega}}\left[A_{1\omega}E_{2}\left(\frac{\widehat{x}}{\text{Kn}_{\omega}}\right) + A_{2\omega}E_{2}\left(\frac{1-\widehat{x}}{\text{Kn}_{\omega}}\right)\right]d\omega. 
\end{equation}
Its Fourier coefficients in Eq.~(\ref{eq:f_expansion}) are then given by:
\begin{equation}
f_0 = \frac{1}{2 \int_0^{\omega_m} \frac{C_{\omega}}{\tau_{\omega}} d\omega} \int_0^{\omega_m}\frac{C_{\omega}\text{Kn}_{\omega}}{\tau_{\omega}}(A_{1\omega}+A_{2\omega})\left[1-2E_3\left(\frac{1}{\text{Kn}_{\omega}}\right)\right]d\omega,
\end{equation}
and
\begin{equation}
f_n =  \frac{1}{ \int_0^{\omega_m} \frac{C_{\omega}}{\tau_{\omega}} d\omega}\int_0^{\omega_m}\int_0^1 \frac{C_{\omega}}{\tau_{\omega}}\text{Kn}_{\omega}\mu\frac{[A_{1\omega}+(-1)^nA_{2\omega}]-e^{-\frac{1}{\text{Kn}_{\omega}\mu}}[(-1)^n A_{1\omega}+A_{2\omega}]}{1+(\text{Kn}_{\omega}\mu)^2(n\pi)^2},
\end{equation}
providing the right-hand side of Eq.~(\ref{eq:LinearMatrix}). Under the same assumption of diffuse, non-black walls, the kernel function becomes
\begin{equation}
K(\widehat{x},\widehat{x}') = \frac{1}{2 \int_0^{\omega_m} \frac{C_{\omega}}{\tau_{\omega}} d\omega}\int_0^{\omega_m} \frac{C_{\omega}G_{\omega}(\widehat{x},\widehat{x}')}{\text{Kn}_{\omega}\tau_{\omega}}d\omega,
\end{equation}
where 
\begin{eqnarray}\nonumber
G_{\omega}(\widehat{x},\widehat{x}') &=&  E_{2}\left(\frac{\widehat{x}}{\text{Kn}_{\omega}}\right)\left[D_{\omega}E_{1}\left(\frac{1-\widehat{x}'}{\text{Kn}_{\omega}}\right)+ B_{1\omega}E_{1}\left(\frac{\widehat{x}'}{\text{Kn}_{\omega}}\right)\right] \\
&+& E_{2}\left(\frac{1-\widehat{x}}{\text{Kn}_{\omega}}\right)\left[D_{\omega}E_{1}\left(\frac{\widehat{x}'}{\text{Kn}_{\omega}}\right)+B_{1\omega}E_{1}\left(\frac{1-\widehat{x}'}{\text{Kn}_{\omega}}\right)\right] + E_{1}\left(\frac{|\widehat{x}-\widehat{x}'|}{\text{Kn}_{\omega}}\right).
\end{eqnarray}
Its Fourier coefficients $k_{mn}$  are given by Eq.~(\ref{eq:k_expansion}), and can be evaluated as: 
\begin{eqnarray}
k_{00} &&= \frac{2}{ \int_0^{\omega_m} \frac{C_{\omega}}{\tau_{\omega}} d\omega}\int_0^{\omega_m}\frac{C_{\omega}\text{Kn}_{\omega}}{\tau_{\omega}}\left\{\frac{2}{\text{Kn}_{\omega}}-1+2E_3\left(\frac{1}{\text{Kn}_{\omega}}\right) \right.  \\ \nonumber
 && \left. + (2D_{\omega}+B_{1\omega}+B_{2\omega})\left[\frac{1}{2}-E_3\left(\frac{1}{\text{Kn}_{\omega}}\right)-\frac{1}{2}E_2\left(\frac{1}{\text{Kn}_{\omega}}\right)+ E_3\left(\frac{1}{\text{Kn}_{\omega}}\right)E_2\left(\frac{1}{\text{Kn}_{\omega}}\right)\right] \right\}d\omega, 
\end{eqnarray}
and
\begin{eqnarray}
k_{m0} &=& \frac{2}{ \int_0^{\omega_m} \frac{C_{\omega}}{\tau_{\omega}} d\omega}\int_0^{\omega_m}\frac{C_{\omega}}{\tau_{\omega}}\int_0^1 \frac{\text{Kn}_{\omega}\mu [(-1)^m+1]\left[e^{-\frac{1}{\text{Kn}_{\omega}\mu}}-1\right]}{1+(\text{Kn}_{\omega}\mu)^2(m\pi)^2}d\mu d\omega\\ \nonumber
&+& \frac{2}{ \int_0^{\omega_m} \frac{C_{\omega}}{\tau_{\omega}} d\omega}\int_0^{\omega_m}\frac{C_{\omega}}{\tau_{\omega}}(D_{\omega}+B_{1\omega})\left[1-E_2\left(\frac{1}{\text{Kn}_{\omega}}\right)\right]\int_0^1 \frac{\text{Kn}_{\omega}\mu\left[1-(-1)^m e^{-\frac{1}{\text{Kn}_{\omega}\mu}}\right]}{1+(\text{Kn}_{\omega}\mu)^2(m\pi)^2}d\mu d\omega \\ \nonumber
&+& \frac{2}{ \int_0^{\omega_m} \frac{C_{\omega}}{\tau_{\omega}} d\omega}\int_0^{\omega_m}\frac{C_{\omega}}{\tau_{\omega}}(D_{\omega}+B_{2\omega})\left[1-E_2\left(\frac{1}{\text{Kn}_{\omega}}\right)\right]\int_0^1 \frac{\text{Kn}_{\omega}\mu\left[(-1)^m -e^{-\frac{1}{\text{Kn}_{\omega}\mu}}\right]}{1+(\text{Kn}_{\omega}\mu)^2(m\pi)^2}d\mu d\omega,
\end{eqnarray}
and 
\begin{eqnarray}
k_{0n} &=& \frac{2}{ \int_0^{\omega_m} \frac{C_{\omega}}{\tau_{\omega}} d\omega}\int_0^{\omega_m}\frac{C_{\omega}}{\tau_{\omega}}\int_0^1 \frac{\text{Kn}_{\omega}\mu [(-1)^n+1]\left[e^{-\frac{1}{\text{Kn}_{\omega}\mu}}-1\right]}{1+(\text{Kn}_{\omega}\mu)^2(n\pi)^2}d\mu d\omega\\ \nonumber
&+& \frac{2}{ \int_0^{\omega_m} \frac{C_{\omega}}{\tau_{\omega}} d\omega}\int_0^{\omega_m}\frac{C_{\omega}\text{Kn}_{\omega}}{\tau_{\omega}}(D_{\omega}+B_{1\omega})\left[\frac{1}{2}-E_3\left(\frac{1}{\text{Kn}_{\omega}}\right)\right]\int_0^1 \frac{\left[1-(-1)^n e^{-\frac{1}{\text{Kn}_{\omega}\mu}}\right]}{1+(\text{Kn}_{\omega}\mu)^2(n\pi)^2}d\mu d\omega \\ \nonumber
&+& \frac{2}{ \int_0^{\omega_m} \frac{C_{\omega}}{\tau_{\omega}} d\omega}\int_0^{\omega_m}\frac{C_{\omega}\text{Kn}_{\omega}}{\tau_{\omega}}(D_{\omega}+B_{2\omega})\left[\frac{1}{2}-E_3\left(\frac{1}{\text{Kn}_{\omega}}\right)\right]\int_0^1 \frac{\left[(-1)^n -e^{-\frac{1}{\text{Kn}_{\omega}\mu}}\right]}{1+(\text{Kn}_{\omega}\mu)^2(n\pi)^2}d\mu d\omega,
\end{eqnarray}
and for $m \neq n$
\begin{eqnarray}
k_{mn} &=& \frac{2}{ \int_0^{\omega_m} \frac{C_{\omega}}{\tau_{\omega}} d\omega}\int_0^{\omega_m}\frac{C_{\omega}}{\tau_{\omega}}\int_0^1 \frac{\text{Kn}_{\omega}\mu \left\{e^{-\frac{1}{\text{Kn}_{\omega}\mu}}[(-1)^m+(-1)^n]-[1+(-1)^{m+n}\right\}}{[1+(\text{Kn}_{\omega}\mu)^2(m\pi)^2][1+(\text{Kn}_{\omega}\mu)^2(n\pi)^2]}d\mu d\omega \\ \nonumber
&+& \frac{2}{ \int_0^{\omega_m} \frac{C_{\omega}}{\tau_{\omega}} d\omega}\int_0^{\omega_m}\frac{C_{\omega}D_{\omega}}{\tau_{\omega}}\int_0^1 \frac{\text{Kn}_{\omega}\mu\left[1-(-1)^m e^{-\frac{1}{\text{Kn}_{\omega}\mu}}\right]}{1+(\text{Kn}_{\omega}\mu)^2(m\pi)^2}d\mu\int_0^1 \frac{\left[(-1)^n -e^{-\frac{1}{\text{Kn}_{\omega}\mu}}\right]}{1+(\text{Kn}_{\omega}\mu)^2(n\pi)^2}d\mu d\omega \\ \nonumber
&+& \frac{2}{ \int_0^{\omega_m} \frac{C_{\omega}}{\tau_{\omega}} d\omega}\int_0^{\omega_m}\frac{C_{\omega}D_{\omega}}{\tau_{\omega}}\int_0^1 \frac{\text{Kn}_{\omega}\mu\left[(-1)^m- e^{-\frac{1}{\text{Kn}_{\omega}\mu}}\right]}{1+(\text{Kn}_{\omega}\mu)^2(m\pi)^2}d\mu\int_0^1 \frac{\left[1-(-1)^n e^{-\frac{1}{\text{Kn}_{\omega}\mu}}\right]}{1+(\text{Kn}_{\omega}\mu)^2(n\pi)^2}d\mu d\omega \\ \nonumber
&+& \frac{2}{ \int_0^{\omega_m} \frac{C_{\omega}}{\tau_{\omega}} d\omega}\int_0^{\omega_m}\frac{C_{\omega}B_{1\omega}}{\tau_{\omega}}\int_0^1 \frac{\text{Kn}_{\omega}\mu\left[1-(-1)^m e^{-\frac{1}{\text{Kn}_{\omega}\mu}}\right]}{1+(\text{Kn}_{\omega}\mu)^2(m\pi)^2}d\mu\int_0^1 \frac{\left[1-(-1)^n e^{-\frac{1}{\text{Kn}_{\omega}\mu}}\right]}{1+(\text{Kn}_{\omega}\mu)^2(n\pi)^2}d\mu d\omega \\ \nonumber
&+& \frac{2}{ \int_0^{\omega_m} \frac{C_{\omega}}{\tau_{\omega}} d\omega}\int_0^{\omega_m}\frac{C_{\omega}B_{2\omega}}{\tau_{\omega}}\int_0^1 \frac{\text{Kn}_{\omega}\mu\left[(-1)^m- e^{-\frac{1}{\text{Kn}_{\omega}\mu}}\right]}{1+(\text{Kn}_{\omega}\mu)^2(m\pi)^2}d\mu\int_0^1 \frac{\left[(-1)^n- e^{-\frac{1}{\text{Kn}_{\omega}\mu}}\right]}{1+(\text{Kn}_{\omega}\mu)^2(n\pi)^2}d\mu d\omega,
\end{eqnarray}
and for $m \neq 0$
\begin{eqnarray}\nonumber
&& k_{mm}=  \frac{2}{ \int_0^{\omega_m} \frac{C_{\omega}}{\tau_{\omega}} d\omega}\left\{\int_0^{\omega_m} \frac{C_{\omega}}{\tau_{\omega}} \frac{\text{tan}^{-1}(m\pi \text{Kn}_{\omega})}{m\pi \text{Kn}_{\omega}}d\omega+2\int_0^{\omega_m}\frac{C_{\omega}}{\tau_{\omega}}\int_0^1 \frac{\text{Kn}_{\omega}\mu \left[e^{-\frac{1}{(-1)^m\text{Kn}_{\omega}\mu}}-1\right]}{[1+(\text{Kn}_{\omega}\mu)^2(m\pi)^2]^2}d\mu d\omega \right\} \\ \nonumber
&&+ \frac{2}{ \int_0^{\omega_m} \frac{C_{\omega}}{\tau_{\omega}} d\omega}\int_0^{\omega_m}\frac{C_{\omega}D_{\omega}}{\tau_{\omega}}\int_0^1 \frac{\text{Kn}_{\omega}\mu\left[1-(-1)^m e^{-\frac{1}{\text{Kn}_{\omega}\mu}}\right]}{1+(\text{Kn}_{\omega}\mu)^2(m\pi)^2}d\mu\int_0^1 \frac{\left[(-1)^m -e^{-\frac{1}{\text{Kn}_{\omega}\mu}}\right]}{1+(\text{Kn}_{\omega}\mu)^2(m\pi)^2}d\mu d\omega \\ \nonumber
&&+ \frac{2}{ \int_0^{\omega_m} \frac{C_{\omega}}{\tau_{\omega}} d\omega}\int_0^{\omega_m}\frac{C_{\omega}D_{\omega}}{\tau_{\omega}}\int_0^1 \frac{\text{Kn}_{\omega}\mu\left[(-1)^m- e^{-\frac{1}{\text{Kn}_{\omega}\mu}}\right]}{1+(\text{Kn}_{\omega}\mu)^2(m\pi)^2}d\mu\int_0^1 \frac{\left[1-(-1)^m e^{-\frac{1}{\text{Kn}_{\omega}\mu}}\right]}{1+(\text{Kn}_{\omega}\mu)^2(m\pi)^2}d\mu d\omega \\ \nonumber
&&+ \frac{2}{ \int_0^{\omega_m} \frac{C_{\omega}}{\tau_{\omega}} d\omega}\int_0^{\omega_m}\frac{C_{\omega}B_{1\omega}}{\tau_{\omega}}\int_0^1 \frac{\text{Kn}_{\omega}\mu\left[1-(-1)^m e^{-\frac{1}{\text{Kn}_{\omega}\mu}}\right]}{1+(\text{Kn}_{\omega}\mu)^2(m\pi)^2}d\mu\int_0^1 \frac{\left[1-(-1)^m e^{-\frac{1}{\text{Kn}_{\omega}\mu}}\right]}{1+(\text{Kn}_{\omega}\mu)^2(m\pi)^2}d\mu d\omega \\ 
&& + \frac{2}{ \int_0^{\omega_m} \frac{C_{\omega}}{\tau_{\omega}} d\omega}\int_0^{\omega_m}\frac{C_{\omega}B_{2\omega}}{\tau_{\omega}}\int_0^1 \frac{\text{Kn}_{\omega}\mu\left[(-1)^m- e^{-\frac{1}{\text{Kn}_{\omega}\mu}}\right]}{1+(\text{Kn}_{\omega}\mu)^2(m\pi)^2}d\mu\int_0^1 \frac{\left[(-1)^m- e^{-\frac{1}{\text{Kn}_{\omega}\mu}}\right]}{1+(\text{Kn}_{\omega}\mu)^2(m\pi)^2}d\mu d\omega.
\end{eqnarray}

These equations specify the matrix elements of $\bar{\bar{A}}$ in Eq.~(\ref{eq:LinearMatrix}). With the linear system specified, the coefficients of the temperature profile $x_m$ can be easily obtained with standard matrix methods. 

\end{document}